\pdfoutput=1 
\documentclass[letter,traditabstract]{aa} 
\usepackage{graphicx}
\usepackage{natbib}
\usepackage{txfonts}
\begin{document}
   \title{The brightest pure-H ultracool white dwarf$^\star$}

   \subtitle{}
\author{S. Catal\'an
       \inst{1},
       P.-E. Tremblay
       \inst{2},
       D. J. Pinfield
       \inst{1},
       L.C. Smith
       \inst{1},
       Z.H. Zhang
       \inst{1},
       R. Napiwotzki
       \inst{1},
       F. Marocco
       \inst{1},
       A.C. Day-Jones
       \inst{3},
       J. Gomes
       \inst{1},
       K.P. Forde
       \inst{1},
       P. W. Lucas
       \inst{1}
       \and
       H.R.A. Jones
       \inst{1}
      }


\offprints{S. Catal\'an, s.catalan@herts.ac.uk \\
$^\star$Based on observations made with ESO telescopes at the Paranal 
Observatory under programme ID 088.C-0048(B).}

\institute{Centre for Astrophysics Research, University of Hertfordshire, Hatfield, AL10 9AB, UK \
  	   \and	     
       Zentrum f\"{u}r Astronomie der Universit\"{a}t Heidelberg, Landessternwarte, K\"{o}nigstuhl 12, 69117 Heidelberg, Germany\
        \and 
	   Departamento de Astronom\'ia, Universidad de Chile, Camino del Observatorio 1515, Santiago, Chile}
\date{\today}

\abstract
{We report the identification of LSR J0745+2627 in the United Kingdom InfraRed Telescope Infrared Deep Sky Survey (UKIDSS) Large Area Survey (LAS) as a cool white dwarf with kinematics and age compatible with the thick-disk/halo population. LSR J0745+2627 has a high proper motion ($890$ mas/yr) and a high reduced proper motion value in the J band ($H_J=21.87$). We show how the infrared-reduced proper motion diagram is useful for selecting a sample of cool white dwarfs with low contamination. LSR J0745+2627 is also detected in the Sloan Digital Sky Survey (SDSS) and the Wide-field Infrared Survey Explorer (WISE). We have spectroscopically confirmed this object as a cool white dwarf using X-Shooter on the Very Large Telescope. A detailed analysis of its spectral energy distribution reveals that its atmosphere is compatible with a pure-H composition model with an effective temperature of $3880\pm90$K. This object is the brightest pure-H ultracool white dwarf ($T_{\rm eff}<4000$K) ever identified. We have constrained the distance (24--45 pc), space velocities and age considering different surface gravities. The results obtained suggest that LSR J0745+2627 belongs to the thick-disk/halo population and is also one of the closest ultracool white dwarfs.}

 


\authorrunning{S. Catal\'an et al.}
\titlerunning{The brightest pure-H ultracool white dwarf.}

\maketitle
%

\section{Introduction}

White dwarfs are the end-product of the evolution of majority of low- and medium-mass stars (with initial masses $<$8 $M_{\odot}$). The evolution of white dwarfs can be expressed as a well-understood cooling process \citep{fon01,sal00}. For this reason white dwarfs are used for cosmochronology, for which they provide an independent age-dating method. The study of the oldest and coolest white dwarfs is particularly interesting because they can provide relevant information about the early star formation of the Galaxy \citep{kil06a}. 


The technique most commonly used to identify cool white dwarfs is the combination of optical and infrared (IR) photometry, together with proper motion  information. Spectroscopic confirmation of cool white dwarfs is also essential, where they should show a featureless spectrum possibly with a weak H$\alpha$ line. Coverage of the IR part of the spectral energy distribution (SED) is very important to distinguish between different compositions: hydrogen, helium or mixed H/He \citep{ber97}. H-rich 
white dwarfs with $T_{\rm eff}<5000$ K are featureless although in the infrared region the presence of molecular hydrogen is evident due to collision induced absorption. Sometimes a pure-He composition is more compatible with the observed photometry, but as pointed out by \cite{ber01}, in the case of white dwarfs with long cooling times this is unlikely, because of accretion from the interstellar medium. According to \cite{kow06} and \cite{kil09}, who computed improved model atmospheres with the Lyman alpha red wing opacity, most of these cool objects are H-rich. 

There are about a dozen ultracool white dwarfs ($T_{\rm eff}<4000$ K) detected in the Sloan Digital Sky Survey \citep[SDSS,][]{gat04,har08} that may be thick-disk or halo WDs, but their SEDs are not reproduced successfully by the current white dwarf model atmospheres, accordingly, their temperatures and ages remain uncertain. More recently, \cite{kil10a} carried out a study of cool white dwarf candidates found in SDSS \citep{har06}. They performed a detailed model atmosphere analysis providing temperatures and compositions, as well as obtaining spectroscopic observations to confirm the white dwarf nature of the candidates. Later, \cite{kil10b} confirmed three of these objects as nearby old halo white dwarf candidates (taking into account the space velocities). Two of these objects, SDSS J213730.87+105041.6 and SDSS J214538.16+110626.6, are cool white dwarfs with H-dominated atmospheres and effective temperatures of 3730-3780 K, being the coolest white dwarfs known in the solar neighbourhood. In a more recent paper, \cite{kil12} obtained parallaxes and confirmed the halo membership of two of the oldest known white dwarfs (SDSS J110217.48+411315.4, hereafter J1102, and WD0346+246). The best-fit pure-H model gives a $T_{\rm eff}$ of  3830 K for the first one, while the latter is compatible with a mixed H/He composition and 3650 K. 


Other investigations cross-matched SDSS and the UKIDSS Large Area Survey \citep[LAS,][]{law07} up to DR6  to identify cool white dwarfs \citep[e.g.~][]{lod09,leg11}, finding several candidates, but all with $T_{\rm eff}>4120$ K. 

In this paper we present our study of LSR J0745+2627. This is the brightest pure-H ultracool white dwarf yet discovered (Sect.~3),  and one of the coolest pure-H white dwarfs known.


\section{LSR J0745+2627 identification}

The object LSR 0745+2627 has previously been identified as a high proper motion object (884mas/yr) by \cite{lep02} and \cite{lep05}. It was listed as a white dwarf candidate based on photometry and reduced proper motion in \cite{rei03}. We re-identified this object as LSR J0745+2627 in the UKIDSS LAS DR9, as a high proper motion object in the proper motion catalogue of Smith et al.~(2012). This catalogue uses LAS DR9 data with a time baseline between two and four years. For this particular object, the J-band image epochs are  2007.1 and 2009.1. For this epoch baseline the proper motion limit is 1 arcsec/yr and the matching radius is 2 arcsec. High proper motion objects can be missed in proper motion catalogues with long epoch baselines because the differences in position are large and matches become harder to identify.

\begin{figure}
\centering
  \includegraphics[width=\columnwidth]{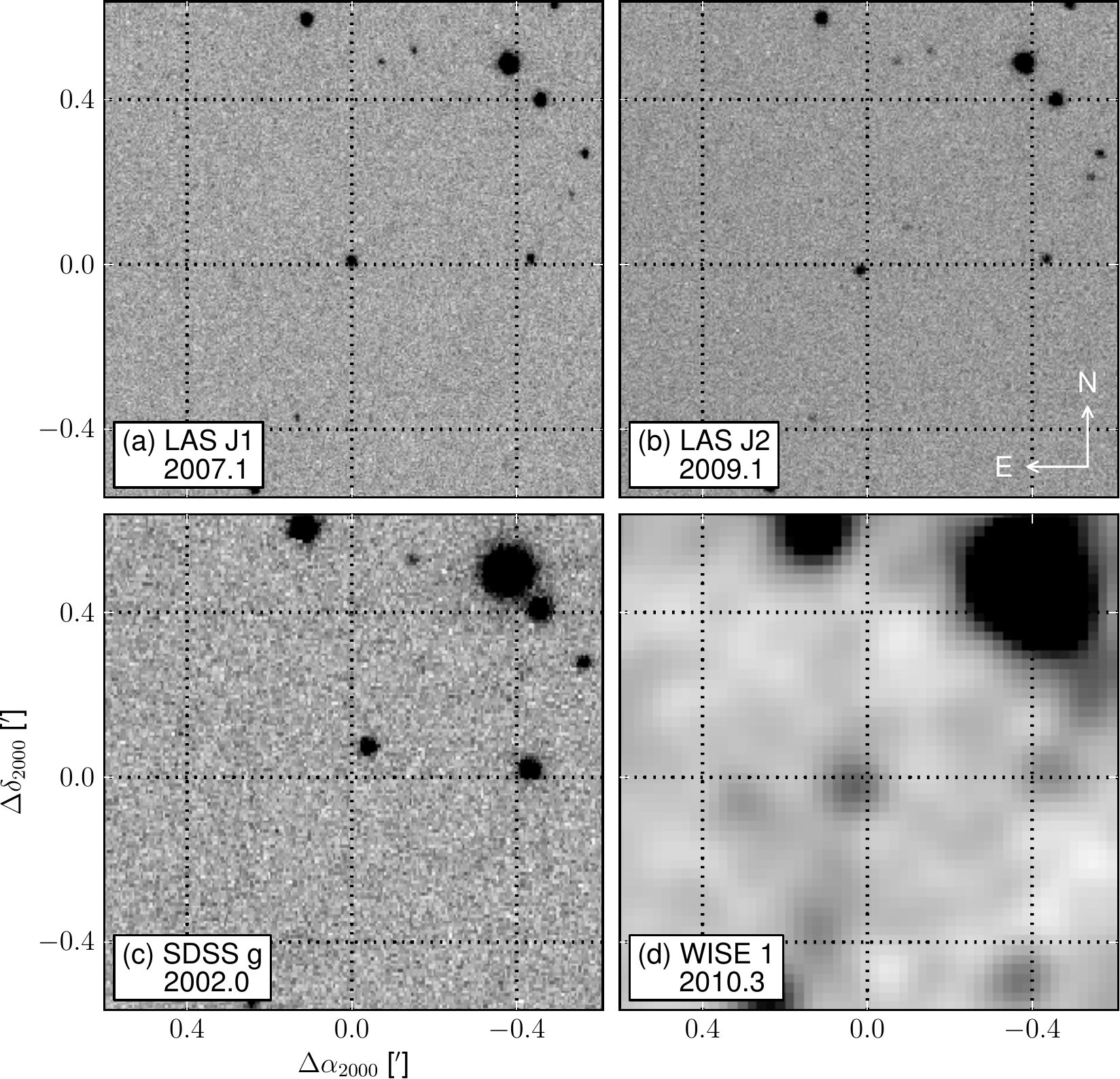}
  \caption{J1 and J2 band LAS images (a and b) and SDSS g band and WISE W1 images (c and d). All images have the same (0,0) point, which corresponds to the LSR J0745+2627 location in the J1 band.}
  \label{fig:ima}
\end{figure}

We obtained photometry from SDSS DR9 and WISE All Sky Release band 1 \citep[W1,][]{wri10} to cover the full spectral range from optical to mid-infrared (see Table \ref{tab:phot}). There was no detection in WISE W2, W3 and W4 at 5$\sigma$ level. A comparison of the photometry of this object with others in the literature reveals that LSR J0745+2627 is the brightest pure-H ultracool white dwarf ever detected (see Sect.~3). The brightest known ultracool white dwarfs are LHS 3250 \citep{har01,ber02} and WD0346+246 \citep{ber01,kil12}, but their SEDs are compatible with He-rich and mixed composition, respectively. J1102 is a pure-H ultracool white dwarf but it is slightly fainter than LSR J0745+2627 \citep{kil12}. Fig.~\ref{fig:ima} shows four $1.2\times1.2$ arcmin images from the LAS (2 J-band epochs, 2007.1 and 2009.1), SDSS (g-band, epoch 2002.0), and W1 (epoch 2010.3). LSR J0745+2627 is the object located at (0,0) in the J1 image. All images have the same (0,0) point. The proper motion information for LSR J0745+2627 is given in Table \ref{tab:phot}. As can be seen, the values obtained by Smith et al. (2012) and \cite{mun04} from USNO-B + SDSS agree within the errors.

\begin{table}
\begin{center}
\caption{Photometric and proper motion information for LSR J0745+2627.} 
\begin{tabular}{lcc}
\hline 
\hline
\noalign{\smallskip}  
Parameter  &  Value\\ 
\noalign{\smallskip}
\hline
\noalign{\smallskip}
UKIDSS LAS DR9 ID  &  +26.4499505+262705.0\\
RA (J2000.0) & 07:45:09.31 \\
DEC (J2000.0) & +26:26:59.82 \\
$Y$ & $17.389\pm0.014$\\
$J$ & $17.115\pm0.013$\\
$H$ & $17.086\pm0.051$\\
$K$ & $17.183\pm0.080$\\ 
$H_J$ & 21.87\\ 
\noalign{\smallskip}
\hline
\noalign{\smallskip}
SDSS DR9 ID & SDSS J074509.02+262705.0\\
RA (J2000.0) & 07:45:09.02\\
DEC (J2000.0) & +26:27:05.01\\
$u$ & 22.65$\pm0.24$\\
$g$ & 19.99$\pm0.02$\\
$r$ & 18.69$\pm0.01$\\
$i$ & 18.23$\pm0.01$\\
$z$ & 17.96$\pm0.02$\\
$H_g$ & 24.74\\ 
\noalign{\smallskip}
\hline
\noalign{\smallskip}
WISE All Sky Release ID & J074509.36+262659.5\\
RA (J2000.0) & 07:45:09.37 \\
DEC (J2000.0) & +26:26:59.54 \\
$W1$ & $16.832\pm0.154$\\
\end{tabular}
\begin{tabular}{lccc}
\noalign{\smallskip}
\hline
\noalign{\smallskip}
\smallskip
$\mu _ \alpha$  $\rm (mas/yr)$ \hspace{1 mm} & $\mu _ \delta$ $\rm (mas/yr)$ \hspace{1 mm} &  Proper motion source\hspace{1 mm} \\ 
$496\pm8$  \hspace{1 mm} &  $-744\pm8$ \hspace{1 mm} &   \cite{lep05} \hspace{1 mm} \\ 
 $500\pm38$  \hspace{1 mm} &  $-706\pm38$ \hspace{1 mm} &  \cite{mun04} \hspace{1 mm} \\
$532\pm11$ \hspace{1 mm}  &  $-716\pm15$ \hspace{1 mm} & Smith et al.~(2012)\hspace{1 mm} \\
\noalign{\smallskip}
\hline
\hline 
\end{tabular}
\label{tab:phot}
\end{center}
\footnotemark[1]{SDSS ugriz magnitudes are on the AB system \citep{fuk96}. LAS YJHK are on the Vega system \citep{hew06}. WISE magnitudes are on the Vega system \citep{wri10}.}\\
\end{table}

\begin{figure}
\centering
  \includegraphics[width=\columnwidth]{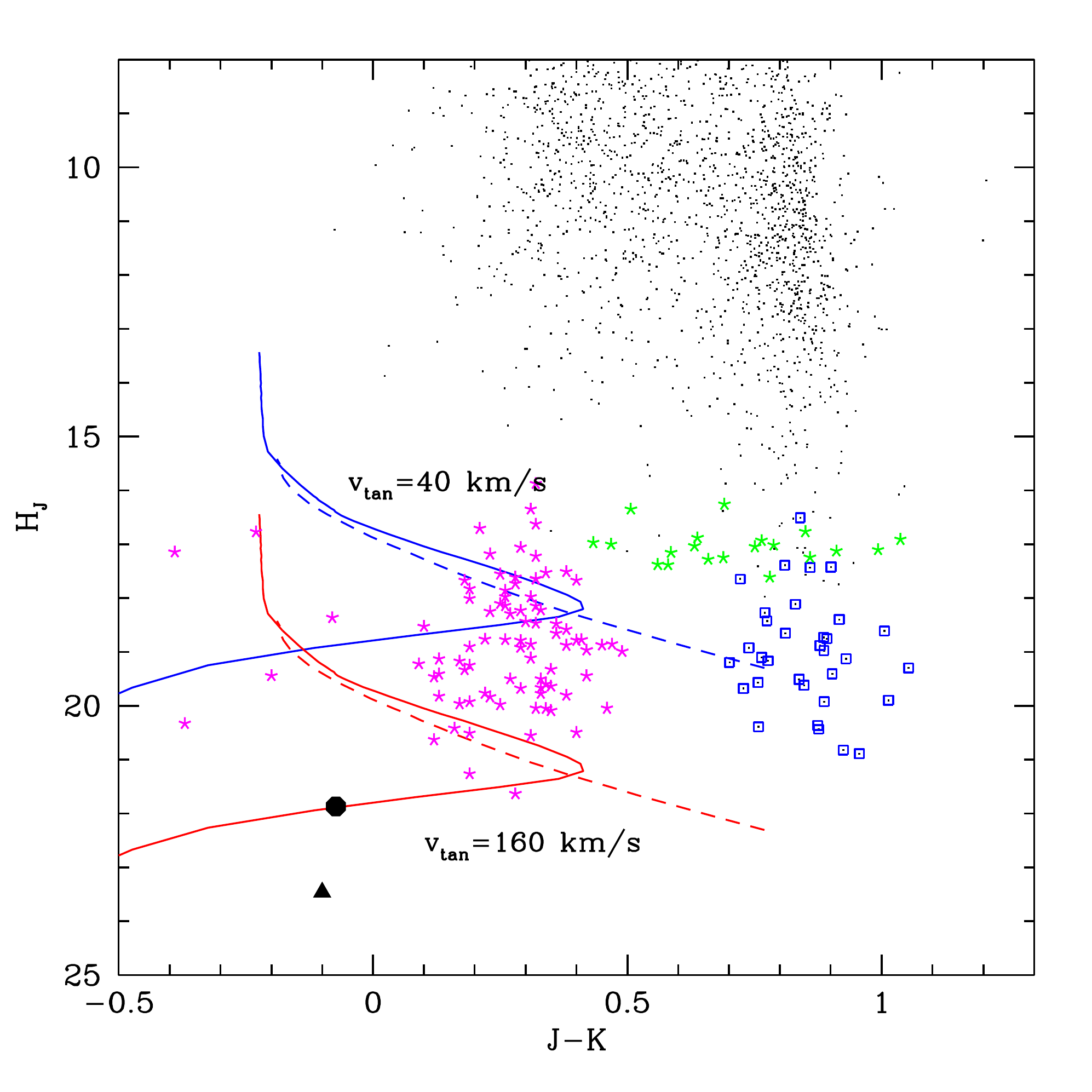}
  \caption{Infrared reduced proper motion diagram. The solid circle corresponds to LSR J0745+2627 and the triangle to the cool white dwarf J1102 \citep{kil12}. Black points correspond to stars from 1 degree around LSR J0745+2627  in SDSS and UKIDSS, and blue squares are stars with pm $>100$ mas/yr. The solid lines correspond to the theoretical tracks for pure-H composition (Sect.~3) for $V_{\rm tan}=40$ km/s (top) and 160 km/s (bottom). The dashed lines correspond to pure-He composition (Sect.~3). Asterisks correspond to objects from \cite{kil10a}: cool white dwarfs are depicted in magenta and subdwarfs in green.}
  \label{fig:rpm1}
\end{figure}

\begin{figure}
\centering
  \includegraphics[width=\columnwidth]{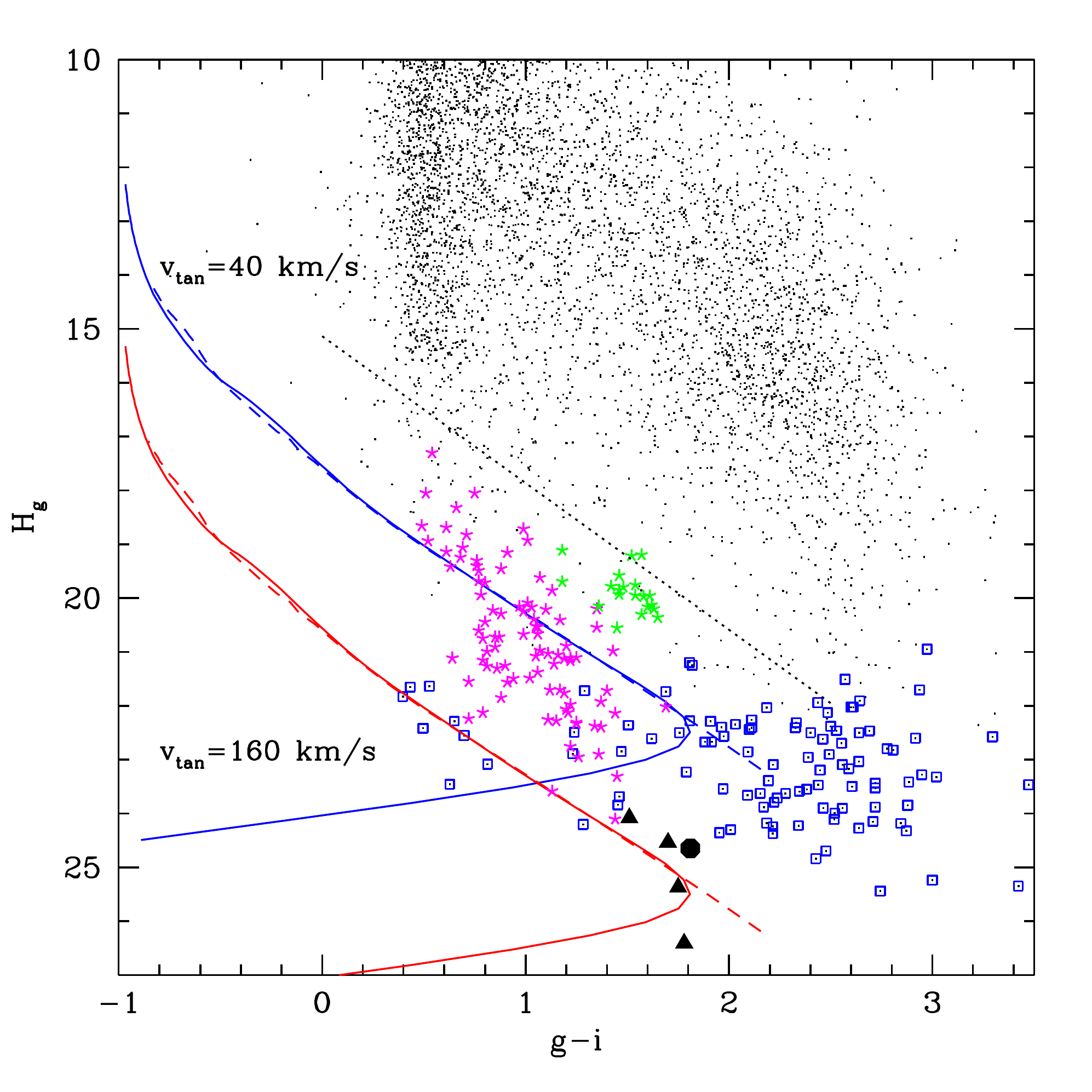}
  \caption{Optical reduced proper motion diagram. The solid circle corresponds to LSR J0745+2627 and the triangles to cool white dwarfs from \cite{kil10b} and \cite{kil12}. Symbols and lines are the same as in Fig.\ref{fig:rpm1}. The region located below the black dotted line in the SDSS diagram represents the white dwarf reduced proper motion cut from \cite{kil06a}.}
  \label{fig:rpm2}
\end{figure}

Proper motion ($\mu$) and reduced proper motion ($H_X=X+5+5\log \mu$, where X is a given magnitude) can be combined with colour information to identify cool white dwarfs. They are also important for distinguishing between different populations (thin, thick-disk and halo) and also for removing contamination from subdwarfs and other objects. In Fig.~\ref{fig:rpm1} we show the J band reduced proper motion diagram that led to the recovery of LSR J0745+2627, which is represented by a solid circle. We also plot stars from SDSS and UKIDSS (that appear in both surveys) located within 1 degree around LSR J0745+2627, to illustrate how useful $H_X$ is to distinguish between normal stars and white dwarfs. Blue squares correspond to high proper motion stars in this sample ($\mu>100$ mas/yr). For the sake of comparison we also show the objects studied in \cite{kil10a}, confirmed cool white dwarfs as magenta asterisks and subdwarfs that contaminated their sample as green asterisks \citep[IR photometry for the subdwarfs is taken from the 2MASS catalogue,][]{cut03}. The solid lines correspond to the theoretical tracks for pure-H composition with tangential velocities of $V_{\rm tan}=40$ km/s (top) and 160 km/s (bottom). The dashed lines correspond to pure-He composition (and the same $V_{\rm tan}$). These tracks correspond to the white dwarf models described in Sect.~3. In the SDSS g band reduced proper motion diagram (Fig.~\ref{fig:rpm2}) LSR J0745+2627 clearly falls within the white dwarf selection criteria $H_g>15.136+2.727(g-i)$ defined by \cite{kil06a}. A comparison between the modelled reduced proper motion and colours suggests a cool H-rich thick-disk/halo white dwarf. LSR J0745+2627 stands out in both diagrams as a blue-to-intermediate colour object with very high reduced proper motion. Although both diagrams are useful for identifying cool white dwarfs, the IR diagram is far more effective since it allows a clear distinction between cool white dwarfs from subdwarfs, which can contaminate optical selections.  

\begin{figure}
\centering
  \includegraphics[width=\columnwidth]{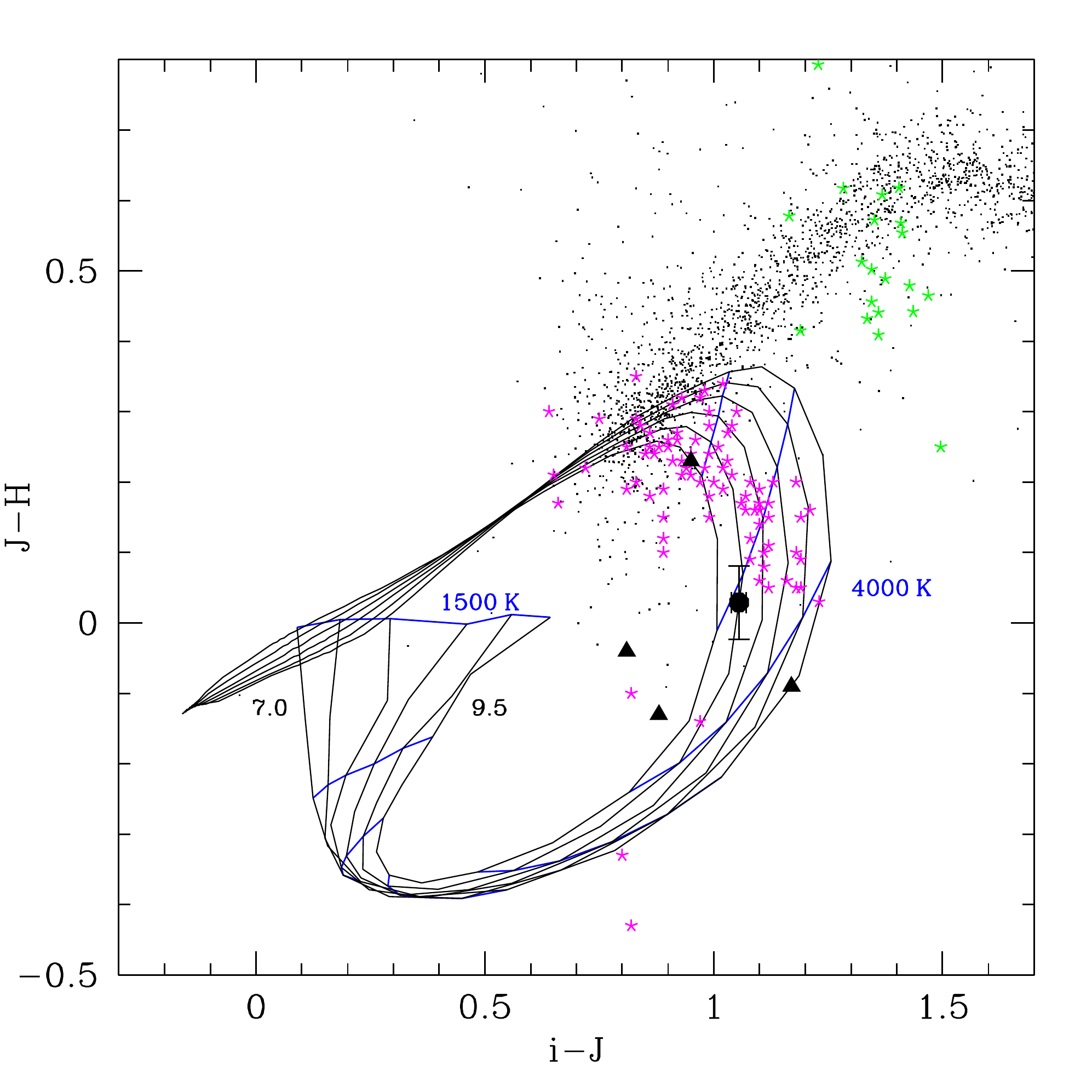}
  \caption{$J-H$ vs. $i-J$ colour-colour diagram. The solid circle corresponds to LSR J0745+2627, and solid triangles to cool white dwarfs from \cite{kil10b} and \cite{kil12}. Black lines correspond to values with equal $\log g$ and blue lines with equal $T_{\rm eff}$ for pure-H composition (Sect.~3). Black points and asterisks represent the same as in Fig.\ref{fig:rpm1}.}
  \label{fig:col}
\end{figure}

Cool white dwarfs also have characteristic colours different from those of other cool stars. In Fig.~\ref{fig:col} we show the $J-H$ vs. $i-J$ diagram with synthetic colours from H-rich model atmospheres overplotted (see Sect.~3).  As can be seen, LSR J0745+2627 (solid circle) is clearly separated from normal stars (black points), and is located close to the 4000 K $T_{\rm eff}$ line. 

It is interesting that this object has not been studied recently as part of targeted SDSS searches. \cite{kil06a} used USNO-B + SDSS proper motions \citep{mun04} to select cool white dwarf candidates in the disk and halo from the $H_g$ vs. $g-i$ diagram. The selection criteria included the requirement that the star had to be detected in all five epochs in USNO-B, however, LSR J0745+2627 is only detected in three plates. In UKIDSS, \cite{lod09} and \cite{leg11} performed a search for cool white dwarfs using SDSS and UKIDSS LAS (DR2 and DR6 respectively). However, LSR J0745+2627 does not appear in the LAS until DR8.

\section{Spectral confirmation and model fitting}

\begin{figure}
\centering
  \includegraphics[width=\columnwidth]{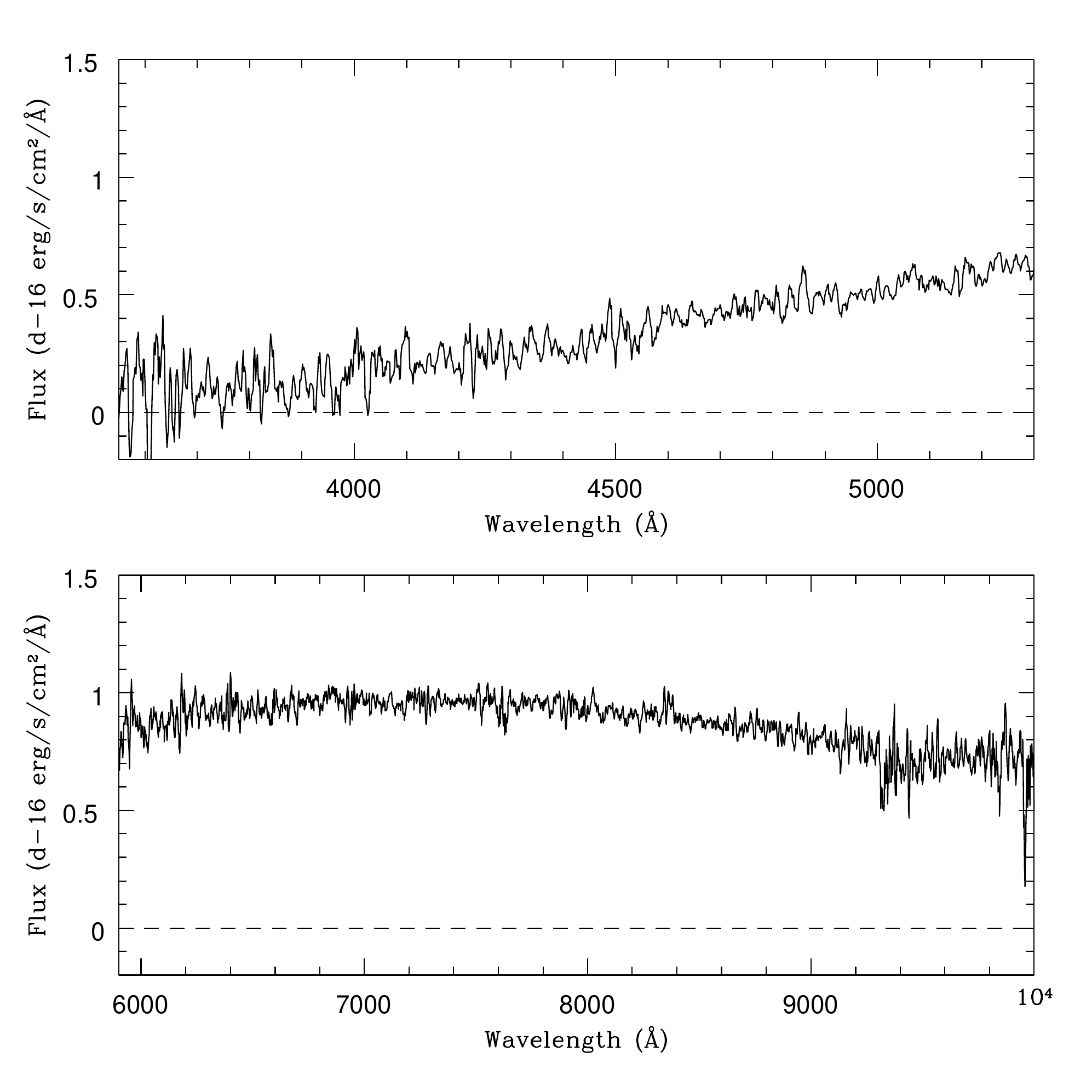}
  \caption{Spectrum obtained at the VLT with a five-pixel boxcar. The top panel corresponds to the UVB arm and the bottom panel to the VIS arm.} 
  \label{fig:spec}
\end{figure}

We obtained a low-resolution spectrum for this object using X-Shooter on the Very Large Telescope (VLT) at Paranal Observatory on 28 January 2012. X-Shooter consists of three arms covering three different wavelength ranges: 3000--5600 $\AA$ for the UVB, 5500--10200 $\AA$ for the VIS and 10200--24800 $\AA$ for the NIR. The instrumental setup used was the echelle mode with a slit width of 1, 0.9 and 0.9 arcsec and with a resolution ($\rm{R}=\lambda/\Delta\lambda$) of 5100, 8800 and 5100 for UVB, VIS and NIR, respectively. The total exposure times are 4x230s in UVB, 4x300s in VIS and 4x390s in NIR.
The signal-to-noise (S/N) obtained for this spectrum is approx. 5--10 in the UVB arm, 10--15 in the VIS arm and below 5 in the NIR. We used the ESO pipeline version 1.3.7 to reduce the data. The star used for the flux calibration was the white dwarf GD 71, while an F star (HD88449) was used for the telluric correction. In Fig.~\ref{fig:spec} we show the spectrum obtained, which is featureless, as expected for cool white dwarfs.
The features at 9400  $\AA$, as well as the two small ones at 6900 $\AA$ and 7600  $\AA$, are residual telluric features that have not been corrected for properly. We do not include the NIR spectrum because it is too noisy and is also affected by two major telluric bands located at 1.35--1.45 and 1.8--1.95 $\mu$m.

\begin{figure*}
\centering
  \includegraphics[width=\columnwidth, angle=-90]{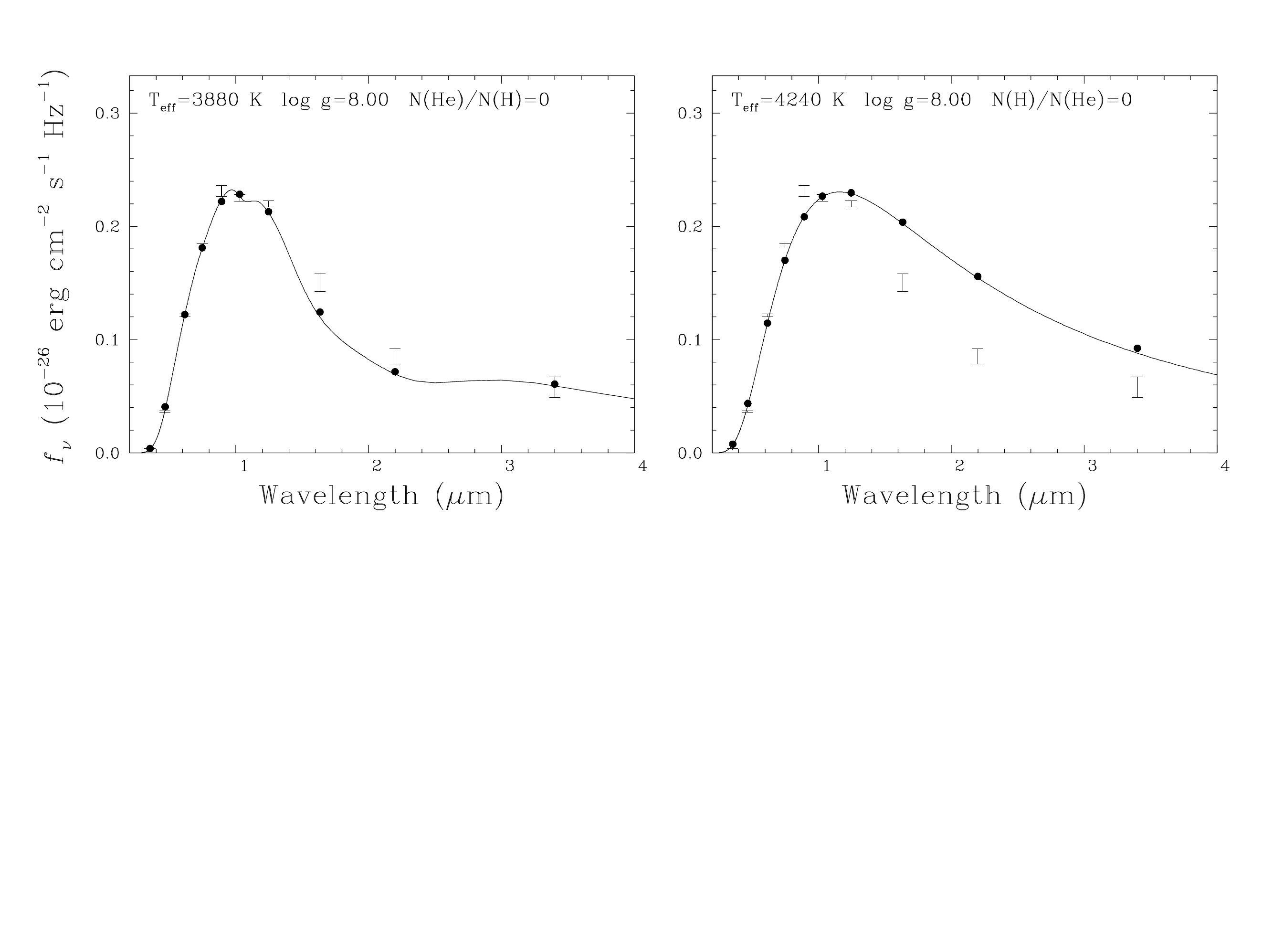}
    \caption{Photometric fit to synthetic magnitudes for a pure-H (left) and pure-He (right) composition. Dots correspond to theoretical fluxes from the models and the errorbars correspond to the SDSS, UKIDSS and WISE photometry.} 
  \label{fig:fit}
\end{figure*}

We performed a fit of the available photometry to synthetic magnitudes considering different compositions to obtain $T_{\rm eff}$. The pure-H models used were those of \cite{tre11}, but now including the red wing opacity from Ly$\alpha$, so the $u$ bandpass is also included \citep{kow06}.  The pure-He model atmospheres considered were the same models as in \cite{kil10a}. The magnitudes were converted into observed fluxes taking into account the appropriate filters \citep{hol06}. Then, the resulting energy distribution was fitted with those derived from model atmosphere calculations, using a non-linear least-squares method. The fitting procedure is explained in detail in \cite{leg11}. The surface gravity was fixed to $\log g=8$ (the most typical value for white dwarfs). As can be seen in Fig.~\ref{fig:fit}, the photometric SED of LSR J0745+2627 is consistent with a very cool pure-H white dwarf (left), while the He-rich composition is not compatible (right). Dots correspond to theoretical fluxes from the models and the error bars correspond to the observed photometry from SDSS, UKIDSS and WISE. The temperature obtained in the H-rich fit is $3880\pm90$ K, which makes this star one of the coolest white dwarfs ever detected. 
A mixed composition was also considered, but adding He to a pure-H composition further enhances the IR-absorption. Since the observed IR-absorption is already weaker than the prediction, the pure-H solution remains the best fit.

\section{Distance, kinematics and membership}

\begin{table}
\begin{center}
\caption{Distances and space velocities for different $\log g$ considered. }
\begin{tabular}{lcccc}
\hline 
\hline
\noalign{\smallskip}  
 $\log g$ $(cgs)$ & 7.5 & 8.0 & 8.5\\ 
\noalign{\smallskip}
\hline
\noalign{\smallskip}      
Distance (pc) & 45 & 34 & 24 \\ 
$V_{\rm tan}$ (km/s) & 183 & 138 & 98 \\ 
$U$, $V$, $W$ (km/s) & 74, -201, 57 & 58, -129, 45 & 44, -86, 34\\
\noalign{\smallskip}
\hline
\hline 
\end{tabular}
\label{tab:vel}
\end{center}
\end{table}

\begin{figure}
\centering
  \includegraphics[width=\columnwidth]{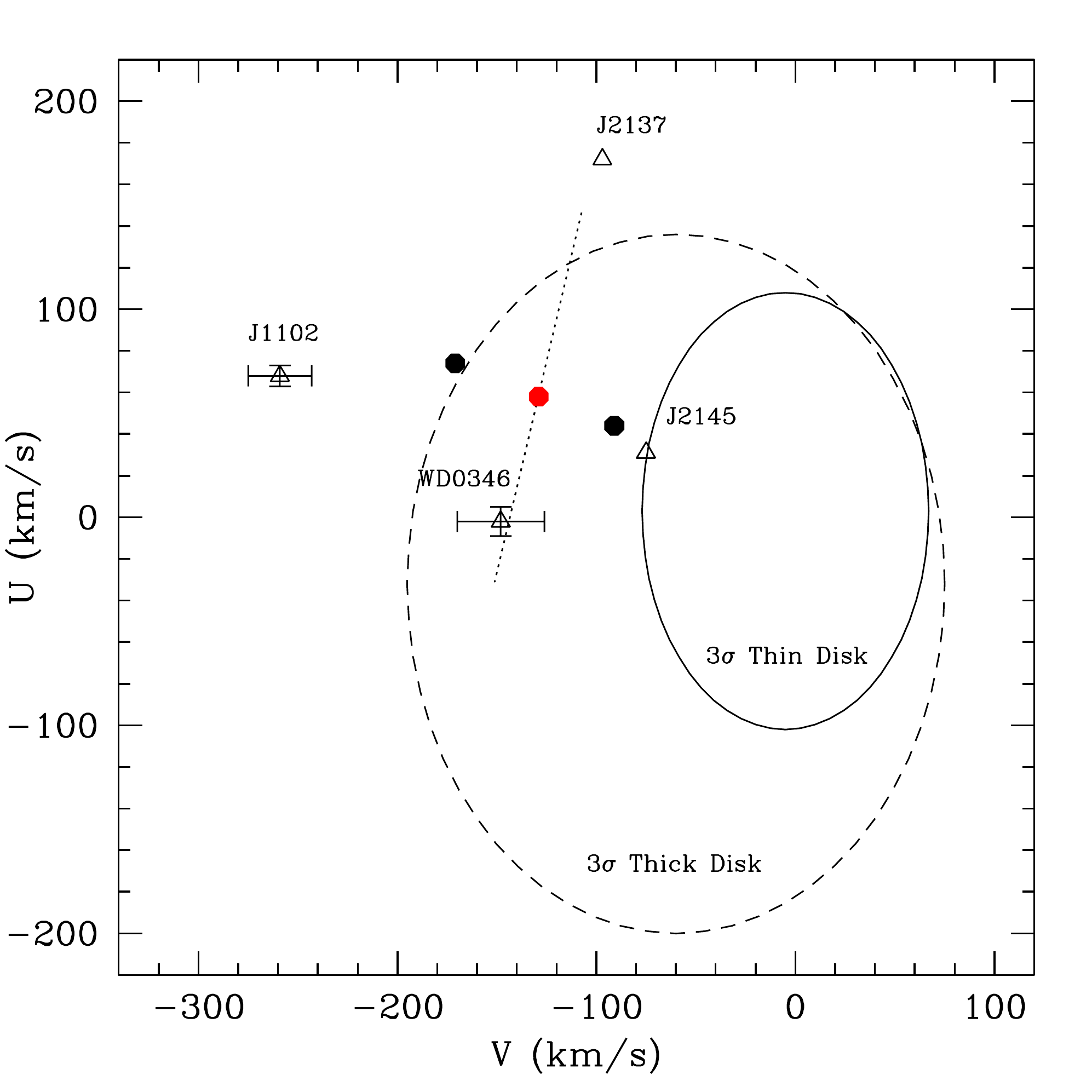}
  \caption{$U$ vs. $V$ space velocities. The circles correspond to LSR J0745+2627 (for $\log g$ range 7.5, 8.0 and 8.5 from left to right) assuming zero radial velocity. The dotted line corresponds to the values considering a radial velocity in the range -100--+100 km/s. Triangles correspond to other known halo/thick-disk cool white dwarfs. We also show the 3$\sigma$ thick-disk and thin-disk ellipses from \cite{pau06}.}
  \label{fig:vel}
\end{figure} 
 
 
Since a parallax measurement is not available for LSR J0745+2627, we determined the photometric distances and tangential velocities considering synthetic absolute photometry for different $\log g$ (see Table \ref{tab:vel}). The tangential velocities range from 98--183 km/s for $\log g$ between 7.5--8.5, i.e., they are compatible with the thick-disk and halo populations. Thus, this object belongs to the local neighbourhood since the distance obtained is in the range 24--45 pc. 
The closest ultracool white dwarfs known are J1102 and WD 0346+246, located at 34 and 28 pc, respectively \citep{kil12}. LSR J0745+2627 could be the closest ultracool white dwarf, but the accurate determination of the mass is essential to confirm this. 
 
We calculated the space velocities following the method described in \cite{pau03} and considering the distances obtained from assuming different $\log g$ (see Table \ref{tab:vel}). In Fig.~\ref{fig:vel} we show a diagram with the space velocities ($U$ vs. $V$). The solid circle corresponds to LSR J0745+2627 considering different $\log g$ (7.5, 8.0 and 8.5) and assuming zero radial velocity. Triangles correspond to halo/thick-disk cool white dwarfs from \cite{kil10b} and \cite{kil12} . The ellipses correspond to the typical values for the thin-disk and the thick-disk (3$\sigma$), taken from \cite{pau06}. The halo ellipse would cover the entire plot and is therefore not represented. If LSR J0745+2627 has a $\log g$ of 8.5, the space velocities would be closer to the thin-disk typical values, but still outside the 3$\sigma$ ellipse. The space velocities better match the kinematics of the thick-disk or halo.

The white dwarf age is another good indication of population membership. Considering the $T_{\rm eff}$ obtained from the SED fitting and using cooling sequences \citep{sal00}, we can obtain the cooling time of a white dwarf. For this we need the mass of the white dwarf. The white dwarf masses corresponding to the same $\log g$ shown in Fig.~\ref{fig:vel} are 0.32, 0.58 and 0.91$M_{\odot}$, and the cooling times are 4.7, 10.6 and 14.7 Gyr, respectively. Accordingly, if LSR J0745+2627 has a high $\log g$, the space velocities could be marginally compatible with the thin-disk, but the cooling time would be much longer than the age of the thin-disk, which rules out this possibility.

The total age of a white dwarf is the sum of the cooling time and the progenitor lifetime in the pre-white dwarf stage. To account for this pre-white dwarf lifetime, we took into account an initial-final mass relationship \citep{cat08} and stellar tracks \citep{dom99}, obtaining progenitor masses $<$0.7 $M_{\odot}$, 1.6 $M_{\odot}$ and 4.3 $M_{\odot}$ and progenitor lifetimes of $>$15 Gyr, 2.2 Gyr and 0.15 Gyr, respectively. Adding these values to the corresponding cooling times, the ages obtained are always compatible with the thick-disk/halo population ($>12$ Gyr).


\section{Conclusions}

We have re-identified LSR J0745+2627 as a high proper motion object in the proper motion catalogue of Smith et al. (2012) using UKIDSS LAS DR9. We studied the location of this object in two reduced proper motion diagrams (J band, g band) and showed that the $H_J$ vs. $J-K$ diagram is very useful for identifying cool white dwarfs and removing contamination from subdwarfs and objects with high proper motion. We  spectroscopically confirmed LSR J0745+2627 as an ultracool ($T_{\rm eff}<4000$K) white dwarf for the first time. The best fit of the SED of this object (including SDSS, UKIDSS and WISE magnitudes) is achieved with pure-H composition model, and the temperature obtained is $3880\pm90$ K. In addition to being one of the coolest pure-H white dwarfs, this object is also the brightest pure-H ultracool white dwarf ever identified. Even though we can not constrain the white dwarf mass due to the lack of parallax measurement, we  constrained the distances, space velocities and total ages considering different surface gravities. The results obtained show that this object is a member of the thick-disk/halo population. The estimated distance of 24--45 pc  makes LSR J0745+2627 one of the closest ultracool white dwarfs ever detected. The parallax determination of this object will help to improve the distance and kinematics, as well as discern between thick-disk and halo membership.

\begin{acknowledgements}
SC acknowledges financial support from the European Commission in the form of a Marie Curie Intra 
European Fellowship (PIEF-GA-2009-237718). P-ET is supported by the Alexander von Huboldt Foundation.
ADJ is supported by a Fondecyt postdoctorado fellowship under project number 3100098. JG is supported by RoPACS, a Marie Curie Initial Training Network funded by the European Commission's Seventh Framework Programme. SC, DP and ADJ have received support from RoPACS during this research. 

\end{acknowledgements}

\bibliographystyle{aa}

\end{document}